\documentstyle[aps]{revtex}
\def\be {\begin{equation}}
\def\ee {\end{equation}}
\def\ba {\begin{eqnarray}}
\def\ea {\end{eqnarray}} 

\def\a  {\alpha}

\def\d  {\delta}

\def\l  {\lambda}

\def\o  {\omega}

\def\p  {\pi}
\def\P  {\Pi}

\def\la {\label}
\def\le {\left}
\def\ri {\right}

\def\f {\frac}
\def\sq {\sqrt}
\def\no {\noindent}
\def\bi {\begin{itemize}}
\def\ei {\end{itemize}}

\def\vs {\vspace}

\def\ul {\underline}

\begin{document}
\draft
\title{Varying Fine Structure Constant and Black Hole Physics}
\author{Saurya Das$^\flat$~,~Gabor Kunstatter$^\sharp$}

\vs{.4cm}

\address{$^\flat$ Department of Mathematics and Statistics, \\ 
University of New Brunswick,\\
Fredericton, New Brunswick - E3B 5A3, CANADA\\
EMail: saurya@math.unb.ca   \\}


\address{$^\sharp$ Dept. of Physics and Winnipeg Institute for Theoretical Physics, \\
The University of Winnipeg, \\
515 Portage Avenue, Winnipeg, Manitoba - R3B 2E9, CANADA \\
EMail: g.kunstatter@uwinnipeg.ca }
\maketitle

\begin{abstract}
Recent astrophysical observations suggest that the value of
fine structure constant
$\alpha=e^2/\hbar c$ may be slowly increasing with time. 
This may be due to 
an increase of $e$ or a decrease of $c$, or both. In this article, 
we argue
from model independent considerations that this 
variation should be considered 
adiabatic. Then, we examine in detail the 
consequences of such an adiabatic variation
in the context of a specific model of quantized charged black holes. 
We find that the second law
of black hole thermodynamics is obeyed, regardless of the origin 
of the variation,
and that interesting constraints arise on the charge and 
mass of black holes.
Finally, we estimate the work done on a black hole of mass $M$ due to the 
proposed $\alpha$ variation.

\end{abstract}

\vs{.6cm}

\section{Introduction}
Some recent astrophysical observations suggest that, $\alpha$,
the fine structure constant of quantum electrodynamics is not a constant, 
but has increased by approximately $\delta\alpha/\alpha \approx 10^{-5}$ 
over $6-10$ billion years, corresponding to a rate of change of about
${\dot\alpha}/\alpha \approx 10^{-16}/year$ \cite{webb,murphy}. 
Although this result is 
unconfirmed and quite controversial (see for example the recent paper by
Bahcall et al \cite{bahcall} where upper bounds on the variation of $\alpha$
and its rate of variation were set at $\delta\alpha/\alpha < 10^{-4}$ over 
about $10$ Billion years and  
${\dot\alpha}/\alpha < 10^{-13}/year$), it provides motivation for 
learning as much as we can about the consequences of such potential 
variations (for other observations see \cite{observe}, 
a review \cite{uzan}, implications of varying $\a$ 
in cosmology \cite{cosmos}, 
in string and brane theory \cite{string} and other 
probable consequences \cite{other}). 

Now, from the definition 
\be
\a = \f{e^2}{\hbar c}
\ee
it is clear that one of the following may be true: 
(i) the electronic charge $e$ is increasing 
(ii) the speed of light $c$ is decreasing 
(iii) Planck's constant $\hbar$ is decreasing. 
Also,
any combination of the above is possible, and moreover, situations such as 
{\it increasing} $\hbar$ and decreasing $c$ (at a faster rate) are also viable.
Although it was suggested by Duff in \cite{duff} that only dimensionless constants 
can legitimately be thought to evolve, here we adopt the viewpoint of Carlip 
\cite{carlip1}, whereby varying $c$, for example, is simply regarded as a 
shorthand for variation of all dimensionless parameters that depend on $c$. 
Moreover,  as will be seen in subsequent sections, our final results depend on the (dimensionless) quantities 
$\a$ and $\d\a/\a$. 

It was argued in \cite{davies} that since the
Bekenstein-Hawking entropy of a  
Reissner-Nordstr\"om black hole is given in terms
of its mass $M$, charge $Q$ and the fundamental constants of nature by:
\be
S_{BH} = \f{k_B \p G}{\hbar c} \le[ M + \sqrt{M^2-\f{Q^2}{G}} \ri]^2~~,
\la{sbh0}
\ee
increase of $Q$ appears to decrease
$S_{BH}$, while decrease of $c$ or $\hbar$ increases $S_{BH}$. Since only the latter is 
consistent with the second law of black hole thermodynamics, the authors of  
\cite{davies} concluded that the variation of $\a$ must be due
to a decrease in $c$ (here and in other articles, for simplicity it 
is assumed that $\hbar$ does not change). 
However, it was recently shown in \cite{carlip2} that the above conclusion
may be somewhat premature. If one rigorously
defines a thermodynamic ensemble by considering a black hole in a `box', then $S_{BH}$ increases whenever $\a$ increases, irrespective of whether
the change is due to an increase of $e$ or decrease of $c$ \cite{carlip2,braden}. 
This means  
that the observed variation of fine structure constant, and the second law of black
hole thermodynamics are perfectly compatible with each other, without 
further assumptions. 
In fact, one might even say that if $\a$ changes at all, it has to increase, 
if black hole thermodynamics has to hold.
The issue of compatibility of varying $\a$ with the second law was also
examined in \cite{youm}. 

In this article, we take a somewhat different approach by first arguing that
on fairly general grounds that the variation of $\alpha$ should be treated
as adiabatic in the context of macroscopic, isolated black holes. For 
concreteness, we then focus on a specific 
model of quantized spherically symmetric charged black
holes, considered in \cite{bdk,dry}, and show that an increase 
in $\a$ automatically implies an increase in $S_{BH}$, once again, 
irrespective of
the source of the variation, and consistent with the conclusions
of \cite{carlip2}. 
We remind the reader that the model considered here as well as the
related spectrum of black hole observables are somewhat speculative. 
However, as was shown in \cite{carlip1}, the qualitative features of
our results may be expected to survive for spectra arising in 
quantum theories of gravity such as string theory and loop quantum gravity.

There are two essential ingredients associated with our analysis, namely:
\begin{itemize}
\item Black holes are isolated dynamical systems that only interact with the 
outside world through the time variation of fundamental constants. 
\item The $\a$ variation is adiabatic (which will be proven in the next section)
\end{itemize}

Under the above assumptions, there are certain quantum numbers that
do not change under the variation. For example, it was first argued
by Bekenstein\cite{bekenstein} that black hole entropy, or horizon area, 
is an adiabatic invariant, and that its quantum spectrum should be
equally space\cite{bekmukh}.
This had been assumed in the algebraic 
approach to black hole area quantization by Bekenstein and Gour \cite{bekgour}. 
Subsequently it was shown in \cite{bdk} that for charged black holes,
it is actually the ``entropy above
extremality'' which is an adiabatic invariant. Thus the latter does not change 
under slow variation of $\a$. However, the total entropy does change and in fact
increases with increasing $\a$. This will be shown in section (\ref{2nd}). 
Another natural consequence of these assumptions is that, 
as with most adiabatic variations in quantum mechanics, 
there will be a shift in energy of the black hole. 
This does not, of course, violate any conservation laws, 
since the energy is provided by `work' being done by external forces
to change the constants.

In the next section, we argue on general grounds that 
for black holes which are far from extremality, 
the observed $\a$-variation can indeed be
regarded as adiabatic.
This result is {\it independent} of any specific model of black hole. 
In subsequent sections we examine the consequences
of the adiabatic invariant in the context of the specific quantum model 
for charged black holes mentioned above. 
Finally, we conclude with a summary and some open questions.

\section{Adiabaticity of Variation of Fine Structure Constant} 
\la{sec2}
First let us consider uncharged black holes, which have 
associated with them a natural time scale
\be
t_{char}= \f{ 2 GM}{c^3}~~.
\label{tchar}
\ee
This scale can either be thought of as the 
characteristic light transition time across one horizon radius, 
or as `inverse-Hawking temperature' $t_{char} = \hbar/4\p k_B T_H$.
Suppose
$H(q,p;\l)$ is the Hamiltonian governing the dynamics of these
black holes with $\l$ any slowly varying parameter. 
In the current context, it could be $\alpha$,
$c$ or $e$. A necessary  condition for adiabaticity is that the parameter
change only a little during the characteristic time scale of the system.
More precisely, \cite{llmech}: 
\be
\f{{\dot \l}}{\l} \ll \f{1}{t_{char}} ~~,
\la{ll1}
\ee
Plugging in the experimental value 
${\dot \a}/\a \approx 10^{-16}/year = 10^{-23}/sec$ \cite{webb}, 
we get from (\ref{tchar}) and (\ref{ll1}) :
$$ \f{c^3}{8\p GM} \gg 10^{-9} ~~,$$
which yields the bound 
\be
M \ll 10^{27} M_\odot~~,
\la{solar1}
\ee
where $M_\odot = 2 \times 10^{33} g$ is the solar mass. 
If the upper bound of Bahcall et al was used instead \cite{bahcall},
namely ${\dot\alpha}/\alpha < 10^{-13}/year$, then the upper bound on
$M$ would have been $10^{24} M_\odot$. In any case, 
since masses of astrophysical black holes are at most of the order of 
billion solar masses, the above bound is always satisfied, and the 
variation of $\a$ can be regarded as adiabatic as far as 
phenomena involving black holes are concerned. 
Adiabatic variations of $\a$ were also considered recently in 
\cite{fair}.

The analysis can easily be extended to black holes with charge, 
in which case, 
\be
t_{char} = \f{\hbar}{4\p k_B T_H} 
= \f{\le[ GM + \sq{(GM)^2 - GQ^2}\ri]^2}{2c^3 \sq{(GM)^2-GQ^2}}
= \f{2GM}{c^3} \le[1 + \f{1}{4} \le( \f{Q}{\sq{G} M}\ri)^4 + \cdots \ri]
\ee
Since charge appears as a fourth-order effect, 
its influence can be ignored for 
all practical purposes in the adiabaticity analysis
(as long as the black hole is away from extremality). 
Similar conclusions would 
follow for black holes with angular momentum, with the replacement
$ Q \rightarrow Jc/\sq{G} M$.
Thus, the bound (\ref{solar1}) appears to be quite robust. 

The above analysis is useful if one knows the adiabatic invariants
associated with black holes. Bekenstein has long argued that the 
entropy of a black hole is an adiabatic invariant \cite{bekenstein}. 
We will now outline a proof that this is so for uncharged black holes.
Our basic assumption is that the time scale $t_{char}$ is due to
an intrinsic periodicity of the system, with 
characteristic frequency:
\be
\o(E) \propto  \f{c^3}{2GM}~~,
\la{oe}
\ee
where $E=Mc^2$. Since black holes are static, it is at first glance difficult
to see what the physical source of this periodicity might be. Two
possibilities have been suggested in the literature. 
This intrinsic periodicity can be connected with
the periodicity in Euclidean time  that characterizes the thermodynamical
properties of the black hole\cite{bdk}, 
in which case the constant of proportionality
is $1/8$. Alternatively, one can use the high damping limit of 
the quasi-normal mode frequency of the black hole
\cite{hod,dreyer,qnm} in which case the
constant of proportionality is $ \ln(3)/4\p$.

It can be shown that for any periodic system, there exists
an adiabatic invariant, which can be calculated (up to a constant shift) as
follows: 
\be
{\cal J} \equiv \f{1}{2\p}~\oint p dq\propto \int \f{dE}{\o(E)} ~~,
\ee
where $(q,p)$ are its phase space variables. Using (\ref{oe}), we get: 
\be
{\cal J} \propto \f{8 GM^2}{c} = \f{\hbar S_{BH}}{2\p k_B}~~,
\la{calj1}
\ee
where $S_{BH}$ is the Bekenstein-Hawking entropy of the Schwarzschild 
black hole. Invoking the Bohr-Sommerfeld quantisation principle yields the result
that black hole entropy is uniformly spaced:
\be
S_{BH} \propto  2 n \p k_B ~~~,~n=0,1,2,\cdots.
\la{calj2}
\ee
From the adiabatic theorem of quantum mechanics, 
under adiabatic variation of any parameter in the entropy, 
the quantum number $n$ does not change \cite{capri}.
We will return to the issue of adiabatic invariance for
charged black holes in the next section.

\section{Charged Black Holes}
\la{2nd}
To further probe the consequences of such adiabatic variation of parameters,
we now consider a specific reduced phase space model of charged black hole
\cite{bdk}. The starting point of this model is the result of \cite{dk,mk} 
that that the dynamics of static spherically symmetric charged configurations in
{\it any} classical theory of gravity in $d$-spacetime
dimensions is governed by an effective action of the form
\be
I = \int dt \le( P_M {\dot M}c^2 + P_Q {\dot Q} - H(M,Q)
\ri) 
\ee
where $M$ and $Q$ are the  mass  and the charge 
respectively and $P_M, P_Q$ the corresponding
conjugate momenta. The momentum $P_M$ has the interpretation of  
asymptotic time difference between the left and right wedges of a Kruskal diagram.
Note that $H$ is independent
of $P_M$ and $P_Q$, such that from Hamilton's
equations, $M$ and $Q$  are constants of motion. 

Now, to incorporate thermodynamics of black holes, one assumes that 
the conjugate momentum $P_M$ is periodic with period equal to  
inverse Hawking temperature (times $\hbar$). That is,
\be
P_M \sim P_M + \f{\hbar}{k_B T_H(M,Q)} ~~.
\la{period}
\ee
Similar assumptions were made in the past using 
different arguments \cite{kastrup}.
Note that the above identification implies that the
$(M,P_M)$ phase subspace has a wedge removed from it,
which makes it difficult, if not impossible to quantise on the
full phase-space.
Thus, one can make a canonical transformation 
$(M,Q,P_M,P_Q) \rightarrow (X,Q,\P_X,\P_Q)$, which
on the one hand `opens up' the phase space, and on the
other hand, 
naturally incorporates the periodicity (\ref{period})
\cite{dk}:
\ba
X &=& \sqrt{\f{\hbar(S_{BH}-S_0(Q))}{\p k_B}} \cos \le( 2\p P_M k_BT_H/\hbar \ri) 
\la{ct1} \\
\Pi_X &=& \sqrt{\f{\hbar(S_{BH}-S_0(Q))}{\p k_B}} \sin \le( 2\p P_M k_BT_H/\hbar \ri) \la{ct2} \\
%
Q &=& Q \la{ct3} \\
\P_Q &=& P_Q + \Phi P_M + S_0'(Q) P_M T_H/k_B   \la{ct4} 
\ea
where the `entropy at extremality' is given by
$S_0(Q) = \pi k_B Q^2/\hbar c$, $'\equiv d/dQ$
and $\Phi$ is the electrostatic potential at the horizon. The
new phase space is ${\cal R}^4$, on which, a rigorous 
quantization can be performed in a straightforward fashion.
Moreover, as shown in \cite{bdk} it is straightforward to calculate
the adiabatic invariant 
for charged black holes in this parameterization. In particular, Eqs.(\ref{calj1}), (\ref{calj2}) generalise to
the following adiabatic invariant :
\be
{\cal J} \equiv \f{1}{2\p} \oint \Pi_X dX 
= \hbar(S_{BH} - S_0(Q))/2\pi k_B = (2n+1)\hbar~~  
\ee
%

Quantization yields  the following spectra for entropy and charge of the
four dimensional
quantum black hole (we refer the reader to \cite{bdk} for details.
For relation of the following spectra with that proposed by 
Bekenstein and collaborators \cite{bekmukh,bekgour}, we refer to \cite{dry}) : 
\ba
S_{BH} &=&  \le(2n + p + 1 \ri) \p k_B~~~~,~~n=0,1,2,\cdots~,
\la{sbh1} \\
\f{Q}{e} &=& m ~~,~m = 0, \pm 1, \pm 2, \cdots 
\la{ch1} \la{fqe1} \\
p &\equiv& \f{Q^2}{\hbar c}= m^2 \alpha
~~~.   \la{ch2}  
\ea
where consistency requires $p$ to be a non-negative integer.
 This in turn implies that the
fine structure constant is constrained to be a rational number:
\be
\a = \f{e^2}{\hbar c} = \f{p}{m^2} ~~, \la{fine1}
\ee
This somewhat strange constraint can be interpreted in either of two ways: 
(a) given an observed value of $\a$, the the 
black hole quantum numbers $p$ and $m$ must be such that it
satisfies this constraint {\it or} (b) even the presence of a single
black hole in the universe constrains the admissible
values of $\a$. We will examine the consequences of this constraint in
the next section, but first we look at the effects of 
a small variation of $\a$
on the entropy(\ref{sbh1})-(\ref{fine1}). 
Since the quantum number $m$ measures the number of constituent
fundamental charged particles making up the black hole, it 
remains fixed during the time in which $\a$ varies.
This implies  $p$ increases if $\a$ increases. Since the quantum number
$n$ is associated with the adiabatic invariant $\cal J$, it remains fixed.
We therefore conclude, via Eq.(\ref{sbh1}), that the
black hole entropy $S_{BH}$ increases monotonically with $\a$, irrespective of 
 the source of the $\a$ variation.
This conclusion is identical to that found in \cite{carlip2}, although we
arrived at it via a different route. While the one hand, we assumed a 
specific model of quantum black holes, on the other hand, we did not
have to use any detailed thermodynamic stability analysis.

\section{Minimum Charge and Mass}
In this section, we show that lower bounds on the mass and charge 
of black holes
follow from requiring the variation of $\a$ to be 
compatible with the constraint (\ref{fine1}).
Consider again Eq.(\ref{fine1}), which implies 
(with $m$ constant as explained before):
\be
\f{\d\a}{\a} = \f{\d p}{p}~~\Rightarrow~~p =\d p~(\d\a/\a)^{-1}~~.
\ee
Now, since $\d\a/\a \approx 10^{-5}$ and $\d p \geq 1$, it follows that
\be
p \geq 10^5~~.
\la{minp}
\ee
In general $p = 10^5\d p$, where $\d p$ is an integer. This, coupled with 
the relations
$Q=\sq{p\hbar c}=\sq{p}~e/\sq{\a}$ and $\a=1/137$
further imply the condition:
\be
 Q \geq  3,000~e \sqrt{\d p} 
\ee
Thus, the minimum charge of a black hole is given by ($\d p=1$):
\be
Q_{Min} = 3,000~e
\la{mincharge}
\ee
and higher charges are square root of integer multiples of the above unit. 
If one uses the limit set by Bahcall et al \cite{bahcall}, namely
$\d\a/a < 10^{-4}$ over about $10$ Billion years, then it can be easily
shown that the bound reduces to $Q_{Min} = 1,000 e$. 

Also, combining (\ref{sbh0}), (\ref{sbh1}) and (\ref{ch1}),  we can write 
the mass of the black hole as:
\be
\f{M}{M_{Pl}} = \f{2n+2p+1}{2\sqrt{2n+p+1}}~~, 
\la{mm1} 
\ee
where $M_{pl}=5 \times 10^{-5} g$ is the Planck mass. 
Let us consider the astrophysically 
relevant, `small charge regime', defined by: 
\be
\f{Q/\sqrt{\hbar c}}{M/M_{Pl}} \equiv k \ll 1~~.
\la{kill}
\ee
Using (\ref{ch2}) and (\ref{mm1}), this gives: 
\be
n = \f{1}{k^2}~\le[p \sqrt{1-k^2} + p~(1-k^2) -\f{k^2}{2} \ri] 
\approx \f{2p}{k^2} \gg p ~~.
\la{nggp}
\ee
In this case, we get from (\ref{mm1}): 
\be
\f{M}{M_{Pl}} = \f{\sqrt{2n+1}}{2}~~,
\la{kill2}
\ee
This, along with (\ref{minp}) and(\ref{nggp}) gives a lower 
bound on the black hole mass:
\be
M_{Min} = 200 M_{Pl}~~.
\la{minmass}
\ee
Again, we recover the lower bound on mass found in \cite{carlip1},
although unlike there, variation of $n$ 
has not been used to derive the result, since the latter does not
in fact change. Once again, the upper bound set by Bahcall et al \cite{bahcall}
would reduce $M_{Min}$ to about $70 M_{Pl}$. 
Also, it may be noted that the lower bounds derived here
hold for charged black holes, leading one to believe that the constraints
(\ref{mincharge}) and (\ref{minmass}) are indeed robust.
We observe from Eqs.(\ref{mincharge}) and (\ref{minmass})
that the minimum $p$ and minimum $n$ both 
increase with decreasing $\delta\alpha/\alpha$.
The situation is similar to one encountered in \cite{carlip1}
for the quantum number $N$.  If future observations were
to make the bound on the $\alpha$ variation even smaller than current
values (as opposed to confirming the results of Webb {\it et al}\cite{webb}),  
the most natural conclusion might be that 
$\delta\alpha=0$, since in this case
$\delta p=0=\delta n$, amounting to no constraints on $p$ and $n$ 
themselves.
%

Finally, let us calculate the work done by a black hole of mass
$M$ due to evolution of $\a$. 
Variation of (\ref{mm1}) yields 
($n$ remaining fixed due to adiabatic invariance). 
\be
\f{\d M}{M} = \f{1}{2}\f{\d c}{c} + \f{6n+2p+3}{2(2n+2p+1)(2n+p+1)}~\d p
~\approx~ \f{1}{2} \f{\d c}{c} + \f{3p}{4n}\f{\d\a }{\a}~\approx~\f{1}{2}\f{\d c}{c}
~~[n \gg p],
\ee
where we have used $M_{Pl} = \sqrt{\hbar c/G} \Rightarrow 
\d M_{Pl}/M_{Pl} = \d c/2c$. 
Since $|\d c/c| \leq \d \a/\a \approx 10^{-5}$, 
the work done due to (so far unknown) external source which
is responsible for the variation of $\a$ is bounded from above by:
\be
W = \d(Mc^2) \leq \f{\d\a}{\a} (Mc^2) = 10^{-5} Mc^2  ~~.
\ee
Thus, for a solar mass black hole, $W \approx 10^{48}~erg$, about
$10^{33}$ times the Planck energy. 

\section{Discussions}

We have argued that the claimed variations of $\a$ should be considered
adiabatic with respect to the dynamics of all physical black holes.
Within a specific model, we showed that such an adiabatic increase of $\a$ predicts an increase of $S_{BH}$, 
no matter what the source of the variation. This conclusion follows from the 
spectra of black hole parameters stated in section (\ref{2nd}).
It differs somewhat from that in \cite{davies}, but
agrees with that in \cite{carlip2}.

However, if the claimed variation of $\alpha$ turns out to be real, 
the  spectra in our model also require minimum values of 
charge and mass, 
roughly equal to $3,000 e$ and $200 M_{Pl}$ respectively. Although
this may seem somewhat disturbing at first, note that these numbers are by 
far much
less than those associated with astrophysical black holes.

There remain, of course many open questions. For example, 
are there other theoretical tests which depend only on variation
of $c$ or $e$, and not both? These could be used 
to determine the source of $\a$ variation \cite{mag}. Is
it legitimate to assume $\hbar$ does not change?
What are the implications of a minimum charge and mass in the spectrum for 
Hawking radiation? Recall that the minimum charge and mass were a direct
consequence of the constraint on the fine structure constant in our model.
Do similar constraints occur in other models of quantum black holes, or can they be avoided? Finally it would be interesting to repeat the above analysis in the context of the spectrum obtained recently by Medved and Gour for Kerr-Newmann black holes\cite{medgour}. 

We hope that answers to at least some of these questions will emerge 
in the near future as more research gets underway in such a fundamental
aspect of experimental and theoretical physics as this. Clearly, 
if the astrophysical observations about $\a$ variation are confirmed by
more precision experiments, it may provide a much needed experimental
laboratory for the study of some features of quantum gravity.

\vs{.4cm}
\no
\ul{{\bf Acknowledgements:}}

\vs{.2cm}
\no
We thank J. Gegenberg and V. Husain for interesting discussions 
and S. Carlip for helpful comments.
This work was supported in part by the Natural Sciences and 
Engineering Research Council of Canada.

\end{document}